\documentclass[conference]{IEEEtran}
\IEEEoverridecommandlockouts
\usepackage{siunitx}
\usepackage{cite}
\usepackage{amsmath,amssymb,amsfonts}
\usepackage{algorithmic}
\usepackage{graphicx}
\usepackage{textcomp}
\usepackage{xcolor}
\def\BibTeX{{\rm B\kern-.05em{\sc i\kern-.025em b}\kern-.08em
    T\kern-.1667em\lower.7ex\hbox{E}\kern-.125emX}}
\begin{document}

\title{MEAN-RIR: Multi-Modal Environment-Aware Network for Robust Room Impulse Response Estimation}
\author{
\IEEEauthorblockN{Jiajian Chen\IEEEauthorrefmark{1}, Jiakang Chen\IEEEauthorrefmark{1}, Hang Chen\IEEEauthorrefmark{1}, Qing Wang\IEEEauthorrefmark{1}\IEEEauthorrefmark{3}, Yu Gao\IEEEauthorrefmark{2}\IEEEauthorrefmark{3}\thanks{\IEEEauthorrefmark{3}Corresponding author}, Jun Du\IEEEauthorrefmark{1}}
\IEEEauthorblockA{\IEEEauthorrefmark{1}University of Science and Technology of China, Hefei 230027, China\\}
\IEEEauthorblockA{\IEEEauthorrefmark{2}AI Research Center, Midea Group (Shanghai) Co.,Ltd., Shanghai 201702, China\\}
\{jjchen1615, jkchen1615\}@mail.ustc.edu.cn,\{hangchen, qingwang2, jundu\}@ustc.edu.cn, gaoyu11@midea.com
}

\maketitle

\begin{abstract}
    This paper presents a Multi-Modal Environment-Aware Network (MEAN-RIR), which uses an encoder-decoder framework to predict room impulse response (RIR) based on multi-level environmental information from audio, visual, and textual sources. Specifically, reverberant speech capturing room acoustic properties serves as the primary input, which is combined with panoramic images and text descriptions as supplementary inputs. Each input is processed by its respective encoder, and the outputs are fed into cross-attention modules to enable effective interaction between different modalities. The MEAN-RIR decoder generates two distinct components: the first component captures the direct sound and early reflections, while the second produces masks that modulate learnable filtered noise to synthesize the late reverberation. These two components are mixed to reconstruct the final RIR. The results show that MEAN-RIR significantly improves RIR estimation, with notable gains in acoustic parameters.
\end{abstract}


\section{Introduction}

Accurate estimation of room impulse responses (RIRs) plays a crucial role in various fields, including virtual and augmented reality \cite{zhang2017surround}, speech enhancement \cite{lebart2001new}, and automatic speech recognition (ASR) \cite{szoke2019building, ratnarajah2020ir}. Traditional RIR measurement methods rely on playing specific signals, which not only incur significant time costs in practical applications but also depend on high-quality playback and reception equipment \cite{guidorzi2015impulse}, making them impractical in many real-world scenarios \cite{7486010}.

For applications such as virtual reality, the acoustic information conveyed by room acoustic parameters like reverberation time alone is not sufficient to generate immersive and realistic spatial audio experiences. These parameters, while useful for analyzing general room acoustics, do not fully capture the complex temporal and spatial characteristics of sound propagation. As a result, a complete RIR must still be estimated to accurately model how sound interacts with the environment. Traditional RIR simulation methods can generally be categorized into wave-based approaches \cite{thompson2006review} and geometric methods \cite{schissler2016interactive}. Wave-based techniques provide accurate physical modeling of sound propagation but are computationally expensive and impractical for real-time applications. On the other hand, geometric approaches, including ray tracing and image source methods, offer efficient approximations but often rely on assumptions that may not hold in all environments. For instance, these methods may neglect the effects of furniture, irregular room shapes, frequency-dependent absorption, and diffraction \cite{chen2024real}, leading to inaccuracy in the simulated RIRs. 

\begin{figure}[t]
  \centering
  \includegraphics[width=\linewidth]{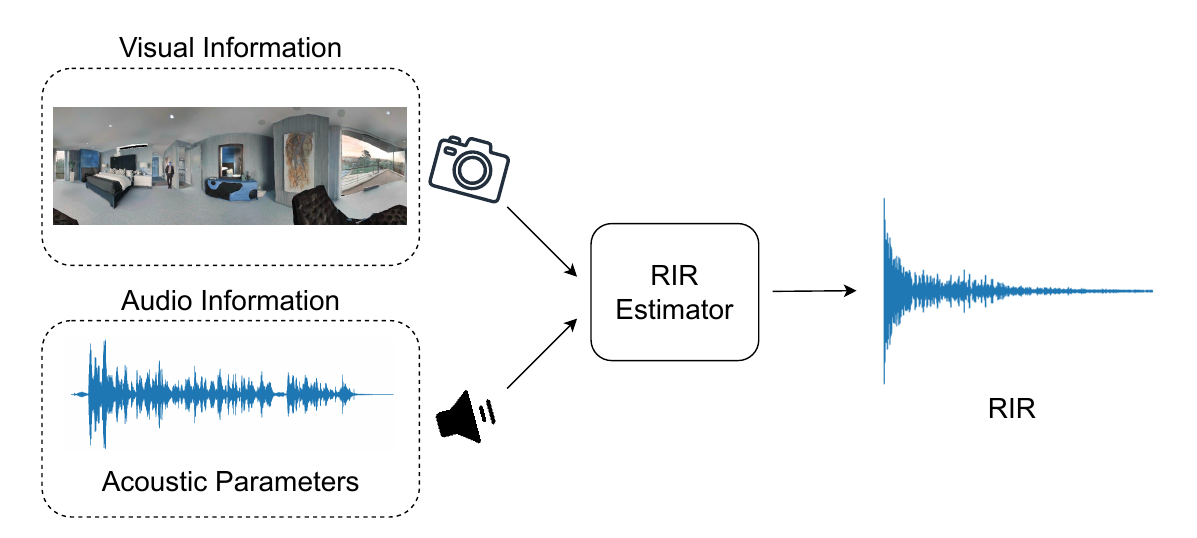}
  \caption{The typical workflow of RIR estimation involves reconstructing the RIR based on available visual and acoustic information.}
  \label{fig:estimation}
\end{figure}

\begin{figure*}[t]
  \centering

  \includegraphics[width=\textwidth]{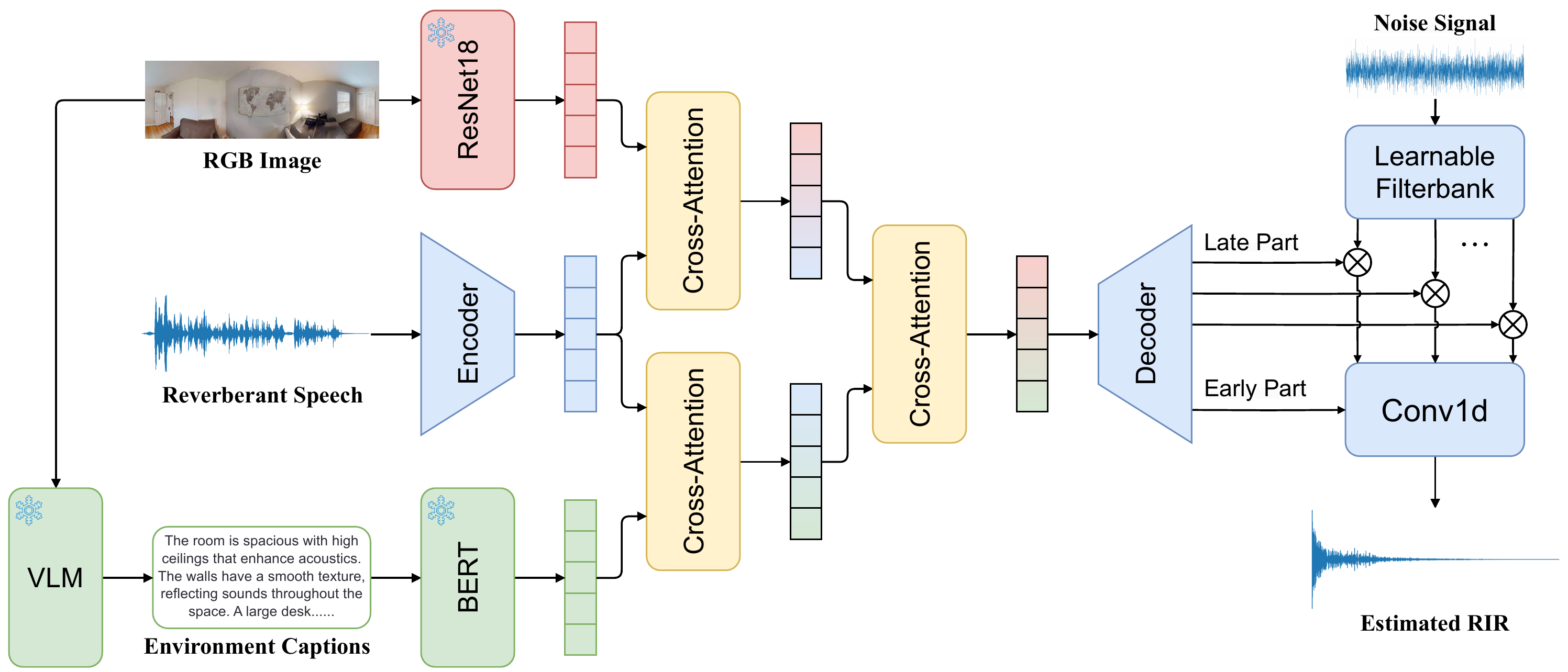}
  \caption{MEAN-RIR consists of three encoders, a multi-modal fusion network, and a decoder with a filtered noise shaping module.}
  \label{fig:structure}
\end{figure*}

With the development of deep learning, researchers have started to explore using neural networks for RIR estimation. For example, acoustic parameters are extracted from captured real-world RIRs and used to generate new synthetic RIRs \cite{ratnarajah2020ir}. Additionally, some studies attempt to directly estimate RIRs from reverberant speech, such as using segmented generative networks to generate sequences of RIR segments \cite{liao2023blind} or employing generative adversarial networks (GANs) to directly estimate RIRs for far-field ASR tasks \cite{ratnarajah2023towards}. Purely visual RIR estimation tasks take a single indoor scene image as input and generate a complete RIR \cite{singh2021image2reverb}. Researchers have also explored using 12 measured RIR samples from one scene along with the room’s geometric layout to estimate the RIR for any speaker-receiver pair in that room \cite{wang2024hearing}. Furthermore, some studies have incorporated visual information, including depth maps and material cues, to improve the estimation accuracy of RIRs, particularly for late reverberation components, through image-to-RIR retrieval techniques \cite{ratnarajah2024av}. An overview of this estimation process is illustrated in Figure~\ref{fig:estimation}, which shows the typical workflow of RIR estimation based on available visual and acoustic information.

Visual modality information has been applied in various speech-related tasks. The RIR estimation task can effectively model the complete acoustic scene using RGB images \cite{majumder2022few}. In dereverberation tasks, scene geometry and audio-visual cross-modal relationships are leveraged to generate an ideal mask for predicting clean speech \cite{chowdhury2023adverb}. Researchers have applied audio-vision multimodal networks to the joint tasks of speech separation and dereverberation \cite{tan2020audio}, aiming to isolate the target speech from background noise, interfering voices, and room reverberation. This approach has led to improved objective intelligibility and perceptual quality of the reconstructed speech. Additionally, textual descriptions of space have proven effective in reconstructing room acoustic scenes. As a supplementary source of spatial information, they can significantly enhance a deep model’s ability to perceive spatial relationships and improve the performance of Visual Text-to-Speech models \cite{he2024multi}.

In this study, we propose a novel multi-modal environment-aware network for robust RIR estimation, termed MEAN-RIR. This method leverages different sources that can provide acoustic environment information, such as reverberant speech, images of the scene, and textual descriptions, to improve RIR estimation performance. Specifically, we treat reverberant speech as the dominant modal and consider panoramic RGB images as auxiliary visual inputs, and textual descriptions obtained from a vision language model (VLM) as supplementary spatial semantic information. MEAN-RIR uses an encoder-decoder architecture, where the outputs of the three encoders are integrated through cross-attention to enable efficient interaction across the different modalities.

The decoder separately estimates the early and late components of the RIR, where the late component is obtained by combining it with a filtered noise shaping network. We conducted experiments on the PanoIR dataset from SoundSpaces 2.0\cite{chen2022soundspaces}. Experimental results demonstrate that the proposed multimodal fusion approach significantly improves RIR estimation performance, with substantial improvement in acoustic parameters such as \(T_{60}\) and DRR.

\section{Approach}

Our network architecture, as illustrated in Figure~\ref{fig:structure}, consists of three encoders, a multi-modal fusion network, and a decoder. The process begins by passing reverberant speech, images, and text through their respective encoders. The outputs of three encoders interacts with each other through cross attention. The decoder employs a filtered noise shaping module to model the late reverberation of the RIR as a sum of noise signals.  

\subsection{Encoder}

Our audio encoder is inspired by the one used in FiNS \cite{9632680}, leveraging 13 time convolution blocks for feature extraction. It adopts a residual structure, utilizing 15×1 large convolutional kernels to capture long-term dependencies while employing 1×1 skip connections to maintain dimensional consistency. The extracted features are then processed through global pooling and a fully connected layer, resulting in a fixed-length 128-dimensional embedding \( \mathbf{A} \in \mathbb{R}^{1 \times 128} \).  

To extract spatial knowledge from RGB images, such as room layouts and object arrangements, we utilize a pre-trained ResNet18 to obtain the image embedding \( \mathbf{V} \in \mathbb{R}^{1 \times 512} \). Additionally, textual descriptions are processed using a pre-trained BERT model \cite{devlin2018bert}, extracting spatial semantic features represented as the text embedding \( \mathbf{T} \in \mathbb{R}^{1 \times 768}  \).  

\subsection{Multi-Modal Fusion Network}

To effectively integrate information from different modalities and enhance the model's spatial perception ability, we have designed a two-step multi-modal fusion network. This network uses three sub-modules with attention mechanisms to perform feature fusion in sequence. Given that the primary modality is audio, we introduce interactions between the audio modality and both the visual and text modalities in each sub-module. The fused audio-textual and audio-visual features are expressed as below:
\begin{equation}
    \mathbf{F_{AT}} = \mathcal{A}(\mathbf{A}, \mathbf{T}, \mathbf{T})
\end{equation}
\begin{equation}
    \mathbf{F_{AV}} = \mathcal{A}(\mathbf{A}, \mathbf{V}, \mathbf{V})
\end{equation}
where \( \mathbf{F_{AT}} \in \mathbb{R}^{1 \times 128} \) and \( \mathbf{F_{AV}} \in \mathbb{R}^{1 \times 128}\) represent the features obtained by the cross-modal attention functions, and \( \mathcal{A}(\cdot) \) denotes the cross-modal attention functions.

Finally, the final fusion step combines the fused features from both modules to obtain the output \( \mathbf{F} \in \mathbb{R}^{1 \times 128} \):
\begin{equation}
    \mathbf{F} = \mathcal{A}(\mathbf{F_{AV}}, \mathbf{F_{AT}}, \mathbf{F_{AT}})
\end{equation}

\subsection{Decoder}

The RIR is typically decomposed into three main components: the direct sound, early reflections, and late reverberation \cite{valimaki2012fifty, valimaki2016more}. For the late reverberation component, an exponential noise model is commonly employed. The FiNS approach leverages this decomposition to estimate the RIR and achieves excellent results by synthesizing the late reverberation signal using noise filtering. We apply this method to the RIR estimation task.

Our decoder is inspired by the generator architecture of GAN-TTS \cite{binkowski2020high}, and is designed to process audio sequences through a series of convolutional blocks with progressive upsampling. Each decoder block consists of two main stages. In the first stage, the input feature is normalized using conditional batch normalization with external condition vectors, then upsampled via a learnable upsample net, and passed through a convolutional layer. A residual connection is introduced using a separate upsampling and projection path to preserve temporal information. The second stage employs dilated convolutions with increasing dilation rates to effectively enlarge the receptive field and capture long-range dependencies in the temporal domain. All convolutional outputs are conditioned using the same auxiliary information, ensuring that the decoder can dynamically adapt its generation behavior based on context. The final decoder output is split into a direct/early component and a late reverberation component, with a sigmoid activation applied to the latter to constrain its amplitude range.

The decoder outputs two components: the early part \(\mathbf{E} \) that denoting an audio clip of 0.05 second duration and a late reverberation mask \(\mathbf{M}\ \in \mathbb{R}\). The early part captures the direct sound and early reflections. Similar to the FiNS approach, we utilize learnable frequency-domain filters to process random noise, with the convolution kernels initialized by octave filters. The filtered noise \(\mathbf{N}\ \in \mathbb{R}\) is then applied to the late reverberation mask and concatenated with the direct sound. Finally, a 1D convolution is used to fuse these features, resulting in the entire RIR. The complete RIR estimation can be expressed as follows:
\begin{equation}
    \hat{\mathcal{RIR}} = \text{Conv1D}(\mathbf{E} \oplus (\mathbf{M} \odot \mathbf{N}))
\end{equation}
where \( \oplus \) denotes concatenation along the time dimension, and
\( \odot \) denotes element-wise multiplication. The output RIR has a fixed length of 44,160 samples, which corresponds to approximately 1 second at a sampling rate of \SI{44.1}{kHz}.

\subsection{Loss Function}

We adopt three different loss functions for training the RIR estimation model: multi-resolution short-term Fourier transformation (STFT) loss, time-domain mean square error (MSE) loss, and energy decay curve (EDC) loss. Let the ground truth RIR as \(R_G\) and the estimated RIR be denoted as \(R_E\).

\noindent\textbf{Multi-resolution STFT Loss} \cite{yamamoto2020parallel}: By combining multiple STFT losses with different parameters, the model can better learn the time-frequency characteristics of speech. The Multi-resolution STFT Loss with 
\(K\) resolutions can be expressed as the average of the sum of two component losses across different resolutions:
\begin{align}
    \mathcal{L}_{\text{SC}} &= \frac{\lVert \,\lvert \text{STFT}(R_G) \rvert - \lvert \text{STFT}(R_E) \rvert \,\rVert_F}{\lVert \lvert \text{STFT}(R_E) \rvert \rVert_F}, \\
    \mathcal{L}_{\text{MAG}} &= \frac{1}{N} \lVert \log \left(\lvert \text{STFT}(R_G) \rvert\right) - \log \left(\lvert \text{STFT}(R_E) \rvert\right) \rVert_1, \\
    \mathcal{L}_{\text{STFT}} &= \sum_{k=1}^{K} \left( \mathcal{L}_{\text{SC}_k} + \mathcal{L}_{\text{MAG}_k} \right).
\end{align}
where \(\lVert\,\cdot\,\rVert_F\) and \(\lVert\,\cdot\,\rVert_1\) denote the Frobenius and \(L_1\) norms, respectively, \(N\) denotes the number of elements in the magnitude.

\noindent\textbf{Time-domain MSE Loss}: For each sample, we calculate the MSE between the estimated RIR and the ground truth RIR:
\begin{align}
    \mathcal{L}_{\text{MSE}} =\sum^{T}_{t=1}(R_G(t)-R_E(t))^2
\end{align}

\noindent\textbf{EDC Loss} : Energy-based loss functions have been shown to be effective in capturing energy-related acoustic characteristics such as \(T_{60}\) and DRR \cite{ratnarajah2023towards}. To further enhance MEAN-RIR's ability to capture these energy-related acoustic features, we introduce an Energy Decay Curve (EDC) loss. The EDC loss is computed by calculating the L2 loss between the energy decay curves of the estimated and ground truth signals:
For each sub-band centered at frequency \(f\), we calculate the energy decay curve as follows:
\begin{align}
    E(R_{G, f}(t))=\sum_{\tau=t}^{T}( R_{G, f}(\tau))^{2}\\
    E(R_{E, f}(t))=\sum_{\tau=t}^{T}( R_{E, f}(\tau))^{2}
\end{align}
where \(R_{G, f}(t)\) and \(R_{E, f}(t)\) denote the sub-band signals of the ground truth and estimated RIR centered at frequency \(f\), respectively. 
The loss function is the total error between the energy decay curves across all frequency bands:
\begin{align}
    \mathcal{L}_{\text{EDC}} = \frac{1}{T}\sum_{f=1}^{F}\sum_{t=1}^T(E(R_{G, f}(t))-E(R_{E, f}(t)))^2
\end{align}
where \(F\) denotes the number of frequency bands. Our final training objective loss is \(\mathcal{L}\), expressed as:
\begin{align}
    \mathcal{L} = \mathcal{L}_{\text{STFT}} + \lambda_1\mathcal{L}_{\text{MSE}} + \lambda_2\mathcal{L}_{\text{EDC}}
\end{align}
where the values of \( \lambda_1 \) and \( \lambda_2 \) are \(10^2\) and \(10^{-1}\), respectively.

\section{Experimental Setup}
\begin{figure}[t]
  \centering
  \includegraphics[width=\linewidth]{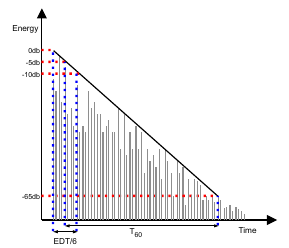}
  \caption{Metrics upon which we evaluate Impulse Response are illustrated.}
  \label{fig:metrics}
\end{figure}

\noindent\textbf{Datasets}: For training and evaluation, We use the PanoIR dataset released by SoundSpaces 2.0 \cite{chen2022soundspaces}. This dataset includes pre-synthesized RIR at a sampling rate of \SI{44.1}{kHz}, RGB images, and depth maps. To obtain spatial semantic information as supplementary knowledge, we used Qwen-VL-Plus \cite{bai2023qwen} to perform text annotations on the panoramic images. We chose scenes from the HM3D dataset for our training, validation, and test sets. Specifically, the training set contains 50 scenes with a total of 50,000 data entries; the validation set includes 6 scenes with 6,000 data entries; and the test set comprises 15 scenes with 2,500 data entries. It is important to note that the scenes in the training, validation, and test sets do not overlap. Due to the lack of a dataset where RIRs are matched with panoramic images in real-world scenarios, we have not tested our method on real data.

For the clean speech, we used \SI{48}{kHz} speech data from the VCTK dataset \cite{veaux2013voice}, which was downsampled to \SI{44.1}{kHz} and used as the dry audio. This dry audio was then convolved with the RIRs from multiple scenes in the PanoIR dataset to generate reverberant speech. Noise has not been considered in this work, but it will be addressed in future research.

The input audio is cropped to 120,000 samples. At the sampling rate of \SI{44.1}{kHz}, this corresponds to approximately 2.7 seconds of audio. The training data consists of approximately 37 hours of audio.

\noindent\textbf{Hyperparameters}: To select the optimal model, we train it using the AdamW optimizer with a batch size of 64 and a learning rate of \(5.5\times10^{-5}\) for 50 epochs. 

\noindent\textbf{Evaluation Metrics}: We quantitatively assess the accuracy of the estimated RIR using standard room acoustic metrics. Reverberation time (\(T_{60}\)), direct-to-reverberant ratio (DRR), and early decay time (EDT) are commonly used room acoustic statistics. These metrics provide essential information about the reverberation characteristics and clarity of sound in a room. \(T_{60}\) measures the time it takes for the sound pressure level to decay \SI{60}{dB}. It reflects the overall sound decay in the space. DRR represents the ratio of sound pressure levels between the direct sound and the reflected sound. This ratio is crucial for evaluating the clarity and intelligibility of the sound. EDT is six times the time it takes for the sound pressure to decay by \SI{10}{dB}. EDT focuses on the early part of the sound decay curve, which affects the clarity and fullness of the sound. Figure~\ref{fig:metrics} shows the metrics used in our study.

\begin{table}[t]
\caption{Performance Comparison of MEAN-RIR with Audio and Audio-Visual Baselines}
\begin{center}
\begin{tabular}{c|c|c|c|c}
\hline
\textbf{Model} & \textbf{Modal} & \textbf{\(T_{60}\) (ms)} & \textbf{DRR (dB)} & \textbf{EDT (ms)} \\
\hline
FiNS\cite{9632680} & A & 76.3 & 3.53 & 162.0 \\
RGB & AV & 65.3 & 2.02 & 87.2 \\
\(*\)AV-RIR \cite{ratnarajah2024av} & AV & 40.2 & 1.76 & 62.1 \\
\textbf{MEAN-RIR} & AVT & \textbf{39.6} & \textbf{1.35} & \textbf{49.9} \\
\hline
\multicolumn{5}{c}{\textit{Ablation Study (AV modality)}} \\
\hline
w/o MSE Loss & AV & 80.8 & 2.43 & 86.4 \\
w/o EDC Loss & AV & 76.9 & 2.01 & 87.2 \\
w/o noise mask & AV & 78.7 & 2.57 & 116.8 \\
w/o cross-attention & AV & 68.5 & 2.18 & 143.2 \\
\hline
\multicolumn{5}{c}{\textit{Ablation Study (AVT modality)}} \\
\hline
w/o MSE Loss & AVT & 51.0 & 1.82 & 72.7 \\
w/o EDC Loss & AVT & 52.0 & 1.46 & 74.1 \\
w/o noise mask & AVT & 54.8 & 1.87 & 65.7 \\
w/o cross-attention & AVT & 43.4 & 1.60 & 98.5 \\
\hline
\end{tabular}
\label{tab:comparison_ieee}
\end{center}
\end{table}

We evaluate the accuracy of the estimated RIR using the Mean Absolute Error (MAE) as the evaluation metric. Specifically, we compute the error between the estimated values and the ground truth for each acoustic parameter mentioned above. 

\noindent\textbf{ASR Evaluation}: We select the AMI dataset\cite{AMI} for ASR evaluation. The AMI Meeting Corpus is a multi-modal dataset consisting of 100 hours of meeting recordings. We use the IHM (Individual Headset Microphone) data as clean speech to synthesize reverberant training data for ASR model training, and use the SDM (Single Distant Microphone) data as the test set.

\section{Results}

\subsection{Baseline and Comparative Models}

The baseline system for comparison is a blind estimation model named FiNS \cite{9632680}, which performs RIR estimation using only reverberant speech. Additionally, we compare with an audio-visual system that uses reverberant speech and RGB image, where visual features are extracted from a pre-trained ResNet18 model. This system is designed to assess the role of textual descriptions as supplementary knowledge. To explore more effective RIR estimation approaches, we conduct various ablation studies to highlight the benefits of different components and loss functions. Specifically, we trained two models by removing the EDC Loss and MSE Loss, respectively, to investigate the impact of these losses on model performance. Furthermore, ablation experiments were conducted to separately validate the effectiveness of using the filtered noise shaping module to simulate late reverberation and the cross-attention mechanism for modality fusion. 

\subsection{Quantitative Results and Analysis}

To evaluate the performance of our proposed method MEAN-RIR, we compare it with AV-RIR \cite{ratnarajah2024av}, a state-of-the-art (SOTA) approach for RIR estimation. AV-RIR is a multi-modal neural network that estimates RIRs from reverberant speech by integrating acoustic features with visual inputs, such as RGB panoramic images and geometric feature maps. Although AV-RIR has shown strong performance, its training data (from SoundSpaces 1.0 \cite{chen2020soundspaces}) and training code are not publicly available. Therefore, we used the training data from SoundSpaces 2.0, which differs in synthesis methodology from SoundSpaces 1.0. Additionally, our test data is not identical to the one used in the AV-RIR paper, which may introduce discrepancies in reported metrics.

Due to the lack of publicly available training scripts, we did not re-train AV-RIR, and instead cite its performance from the original paper. All other baseline models were trained and evaluated on the same dataset as MEAN-RIR to ensure fair comparison. While differences in data and experimental setups must be taken into account, we believe the comparison remains informative and provides useful reference points for evaluating RIR estimation methods. 
We did not include a comparison with S2IR-GAN\cite{ratnarajah2023towards} in our study, as its training scripts, along with the associated training and testing datasets, have not been publicly released. Except for the results of AV-RIR, all other results are obtained using the same training and testing data.

\begin{figure}[t]
  \centering
  \includegraphics[width=\linewidth]{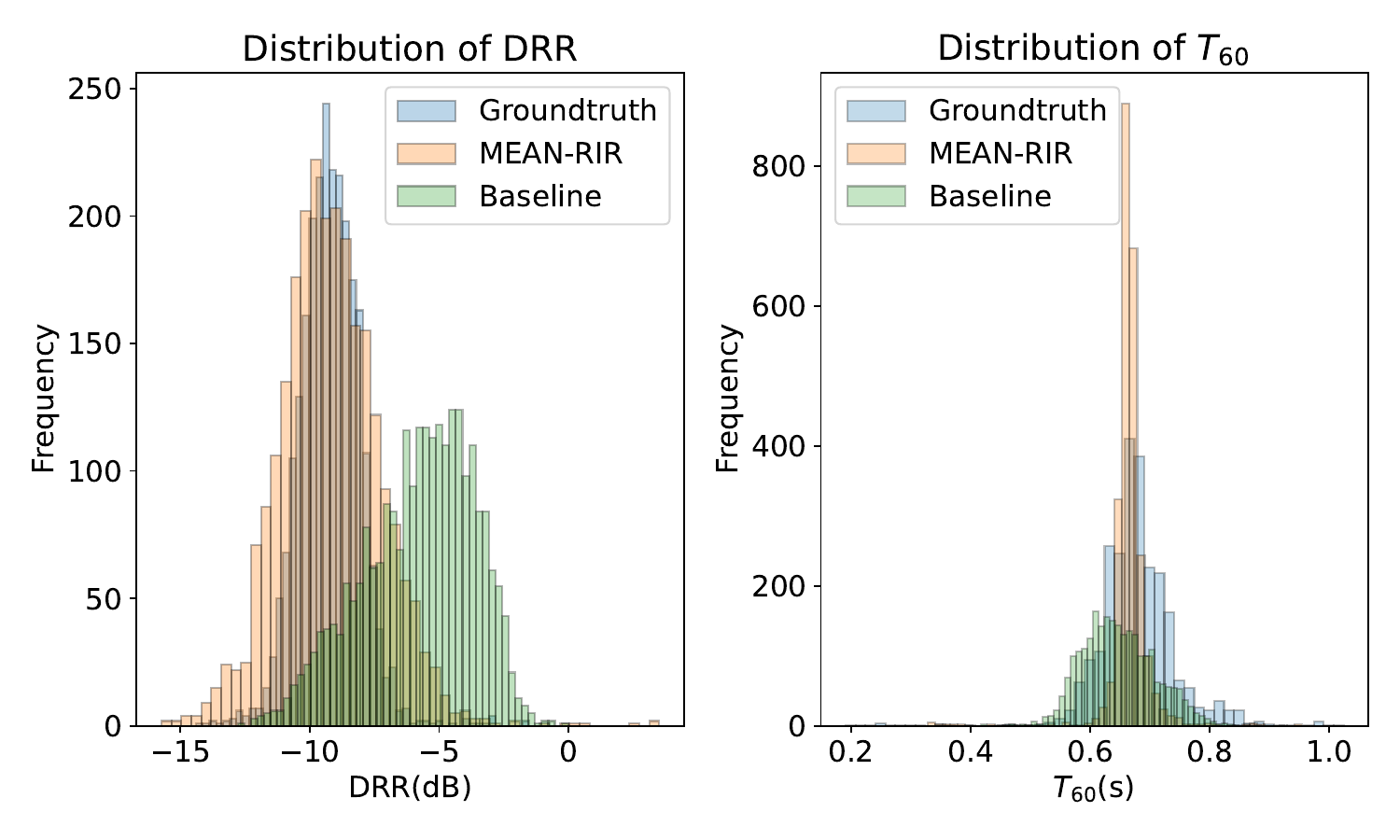}
  \caption{DRR and \(T_{60}\) distributions of ground truth RIR, RIR estimated by MEAN-RIR, and the baseline model are presented. These distributions are compared across the same dataset to evaluate the performance and accuracy of the different methods in estimating these acoustic parameters.}
  \label{fig:distribution}
\end{figure}

Table~\ref{tab:comparison_ieee} shows the performance of different models in terms of \(T_{60}\), DRR, and EDT. The model trained without MSE loss performs poorly on DRR. This is primarily because the MSE loss helps the model capture the RIR waveform more accurately, especially in terms of time-domain errors. Without the MSE loss, the model may fail to accurately model the early components of the RIR, which is crucial for estimating the energy difference between direct sound and reverberation. The EDC loss has a clear positive effect on improving \(T_{60}\). This is likely because it aids the model in better learning the energy decay characteristics. The introduction of the noise mask module allows the model to more accurately model the late reverberation components, which also leads to an improvement in DRR. Surprisingly, the model that directly concatenates features performs better than expected on \(T_{60}\) and DRR, but struggles to learn the accurate direct sound information required for EDT. This suggests that the cross-attention mechanism plays a significant role in improving the model’s performance.

\subsection{Distribution Analysis of Acoustic Metrics}

We plot a comparison of the DRR distributions for intuitive analysis. Figure~\ref{fig:distribution} compares the DRR distributions of the ground truth RIR, the RIR estimated by our MEAN-RIR model, and the baseline model. The distribution of the ground truth RIR is relatively narrow and concentrated, indicating a smaller range of DRR values, which suggests that the true RIR has less variation in terms of the direct-to-reverberant ratio. In contrast, the DRR distribution of the RIR estimated by MEAN-RIR is broader, reflecting the model's ability to account for a wider range of reverberation characteristics, while the baseline model’s RIR estimation shows greater deviation. The results highlight the effectiveness of MEAN-RIR in capturing the key features of DRR, demonstrating its ability to approximate the true reverberation properties with reasonable accuracy. Additionally, we also compare the \(T_{60}\) distributions, as shown in Figure~\ref{fig:distribution}. The \(T_{60}\) distribution of the RIR estimated by MEAN-RIR is more concentrated, indicating that MEAN-RIR tends to produce more consistent reverberation times. This concentration suggests that MEAN-RIR is better at capturing the expected reverberation characteristics.

\begin{figure}[t]
  \centering
  \includegraphics[width=\linewidth]{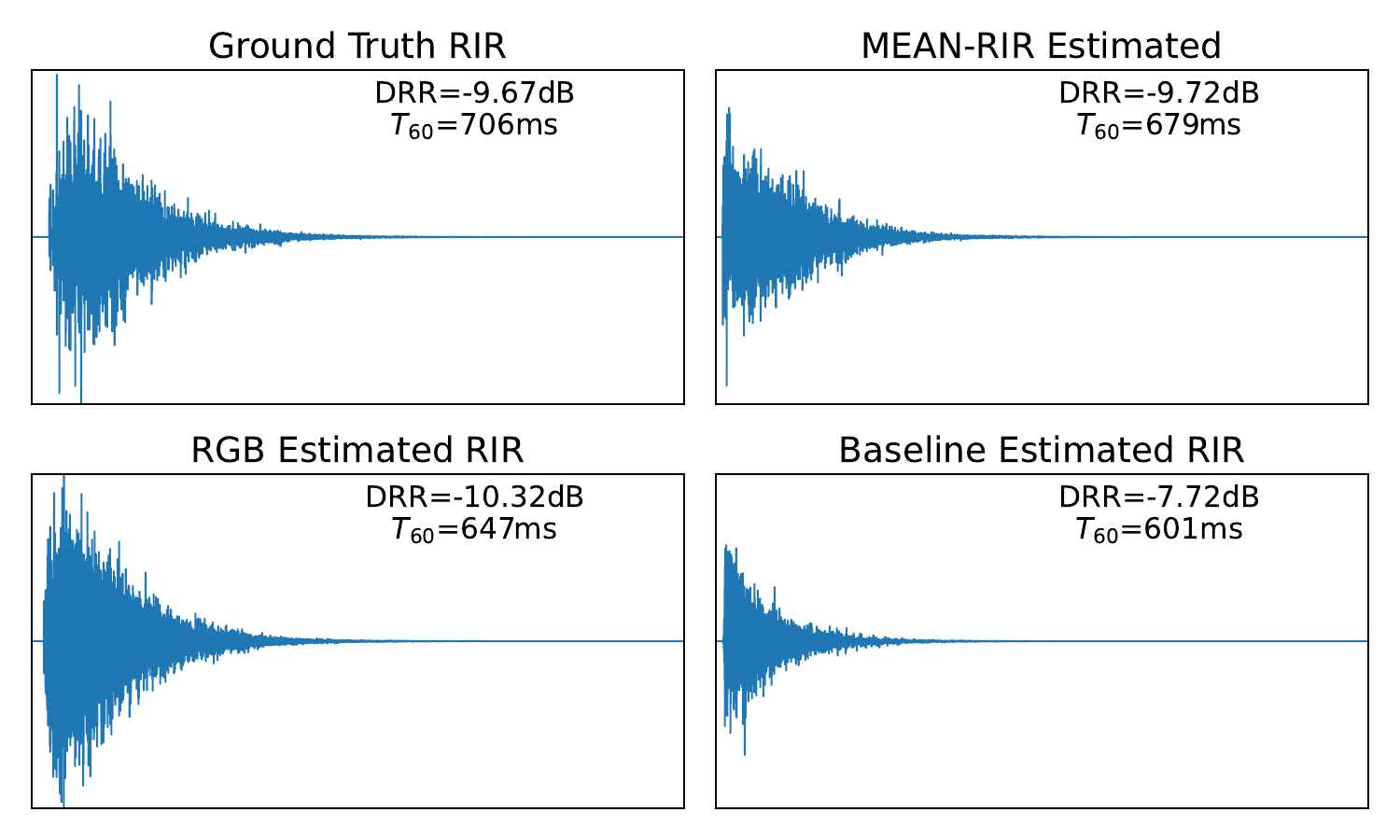}
  \caption{Comparison of the RIR waveform estimated by MEAN-RIR with the Ground Truth and other models.}
  \label{fig:wav}
\end{figure}

\subsection{Time-Domain RIR Comparison}

In Figure~\ref{fig:wav}, we present the RIR waveforms estimated by MEAN-RIR, compared with the ground truth RIR waveform, along with the waveforms of an audio-only baseline model and an audio-visual model. It is evident that MEAN-RIR achieves the best time-domain waveform accuracy. In terms of acoustic parameter estimation, it outperforms the other two models due to the effective fusion of multi-modal information. This demonstrates its effectiveness in 
RIR estimation and generating an RIR perceptually perceptually close to the ground truth.

\subsection{ASR Evaluation}

We comprehensively evaluate the effectiveness of our proposed MEAN-RIR method within the framework of an ASR task using the WeNet toolkit. Specifically, we conduct experiments utilizing the AMI corpus\cite{AMI}, which provides both close-talk (IHM) and far-field (SDM) speech recordings. The close-talk IHM data represents clean speech recorded via headset microphones, while the SDM data simulates realistic reverberant conditions with microphone arrays placed at a distance.

Our approach utilizes the previously introduced PanoIR-synthesized test data rather than the SDM recordings, as the MEAN-RIR model requires panoramic visual information as input and thus cannot be applied directly to SDM speech. Specifically, we use MEAN-RIR to estimate RIRs based on the available panoramic images and their corresponding reverberant speech. These estimated RIRs are then convolved with the clean IHM speech to generate new training data that simulates realistic reverberant conditions. This strategy enables the creation of a large-scale dataset with reverberation characteristics closely resembling real-world far-field environments.

Using this synthetic reverberant training data, we train an ASR model based on WeNet 2.0\cite{wenet2.0}. To assess the generalization performance of the model, we evaluate it on the real SDM test set, which contains naturally reverberated speech. As a baseline for comparison, we also train another ASR model using only the original clean IHM data without any reverberation augmentation.

\begin{table}[t]
\caption{ Far-field ASR results were obtained for far field speech data recorded by single distance microphones (SDM) in the AMI corpus. The best results are shown in bold}
\begin{center}
\begin{tabular}{l|c}
\hline
\textbf{Training Dateset} & Word Error Rate  \\
Clean Speach $\circledast$ RIR &  [\%] \\
\hline
IHM $\circledast$ None & 58.7 \\
\hline
\textbf{IHM} $\circledast$ \textbf{MEAN-RIR (ours)} & \textbf{54.8}  \\
\hline

\end{tabular}
\label{tab:asrresult}
\end{center}
\end{table}

We use the word error rate (WER) as the primary evaluation metric. The results are presented in Table~\ref{tab:asrresult}. A lower WER reflects better ASR performance. The results comparison indicates that the reverberation synthesized using the MEAN-RIR method effectively mimics real acoustic environments. This demonstrates the potential of MEAN-RIR to enhance ASR robustness under mismatched training and testing conditions caused by reverberation.

\section{Conclusions}

This paper presents MEAN-RIR, a deep learning framework for RIR estimation that utilizes multi-modal inputs, including reverberated speech, environmental images, and textual descriptions. Cross-attention mechanism and noise-filtering decoder enhance the model’s ability to capture environmental context and accurately estimate late RIR components. Experimental results show MEAN-RIR outperforms existing methods, with significant improvements in key acoustic parameters including reverberation time, direct-to-reverberant ratio, and early decay time. These results demonstrate the potential of multi-modal inputs for improving RIR estimation. Future work will focus on optimizing multi-modal fusion and exploring its application in various acoustic environments.

\section*{Acknowledgment}

This work was supported by the National Natural Science Foundation of China under Grant No. 62171427.

\vspace{12pt}
\color{red}


\begin{thebibliography}{00}
\bibitem{zhang2017surround} W. Zhang, P.N. Samarasinghe, H. Chen, and T.D. Abhayapala, ``Surround by sound: a review of spatial audio recording and reproduction,'' *Applied Sciences*, vol. 7, no. 5, pp. 532, 2017.
\bibitem{lebart2001new} K. Lebart, J. Boucher, and P.N. Denbigh, ``A new method based on spectral subtraction for speech dereverberation,'' *Acta Acustica united with Acustica*, vol. 87, no. 3, pp. 359--366, 2001.
\bibitem{szoke2019building} I. Sz{\"o}ke, M. Sk{\'a}cel, L. Mo{\v{s}}ner, J. Paliesek, and J. {\v{C}}ernock{\`y}, ``Building and evaluation of a real room impulse response dataset,'' *IEEE Journal of Selected Topics in Signal Processing*, vol. 13, no. 4, pp. 863--876, 2019.

\bibitem{ratnarajah2020ir} A. Ratnarajah, Z. Tang, and D. Manocha, ``{IR-GAN}: room impulse response generator for far-field speech recognition,'' *Proc. Interspeech 2021*, pp. 286--290, 2021.
\bibitem{guidorzi2015impulse} P. Guidorzi, L. Barbaresi, D. D’Orazio, and M. Garai, ``Impulse responses measured with {MLS} or {Swept-Sine} signals applied to architectural acoustics: an in-depth analysis of the two methods and some case studies of measurements inside theaters,'' *Energy Procedia*, vol. 78, pp. 1611--1616, 2015.
\bibitem{7486010} J. Eaton, N.D. Gaubitch, A.H. Moore, and P.A. Naylor, ``Estimation of room acoustic parameters: the {ACE Challenge},'' *IEEE/ACM Transactions on Audio, Speech, and Language Processing*, vol. 24, no. 10, pp. 1681-1693, 2016.
\bibitem{thompson2006review} L.L. Thompson, ``A review of finite-element methods for time-harmonic acoustics,'' *The Journal of the Acoustical Society of 
America*, vol. 119, no. 3, pp. 1315--1330, 2006.
\bibitem{schissler2016interactive} C. Schissler and D. Manocha, ``Interactive sound propagation and rendering for large multi-source scenes,'' *ACM Transactions on Graphics (TOG)*, vol. 36, no. 4, pp. 1, 2016.
\bibitem{chen2024real} Z. Chen, I.D. Gebru, C. Richardt, A. Kumar, W. Laney, A. Owens, and A. Richard, ``{Real Acoustic Fields}: an audio-visual room acoustics dataset and benchmark,'' *Proceedings of the IEEE/CVF Conference on Computer Vision and Pattern Recognition*, pp. 21886--21896, 2024.
\bibitem{liao2023blind} Z. Liao, F. Xiong, J. Luo, M. Cai, E.S. Chng, J. Feng, and X. Zhong, ``Blind estimation of room impulse response from monaural reverberant speech with segmental generative neural network,'' **, 2023.
\bibitem{ratnarajah2023towards} A. Ratnarajah, I. Ananthabhotla, V.K. Ithapu, P. Hoffmann, D. Manocha, and P. Calamia, ``Towards improved room impulse response estimation for speech recognition,'' *ICASSP 2023-2023 IEEE International Conference on Acoustics, Speech and Signal Processing (ICASSP)*, pp. 1--5, 2023.
\bibitem{singh2021image2reverb} N. Singh, J. Mentch, J. Ng, M. Beveridge, and I. Drori, ``{Image2Reverb}: cross-modal reverb impulse response synthesis,'' *Proceedings of the IEEE/CVF International Conference on Computer Vision*, pp. 286--295, 2021.
\bibitem{wang2024hearing} M.L. Wang, R. Sawata, S. Clarke, R. Gao, S. Wu, and J. Wu, ``Hearing anything anywhere,'' *Proceedings of the IEEE/CVF Conference on Computer Vision and Pattern Recognition*, pp. 11790--11799, 2024.
\bibitem{ratnarajah2024av} A. Ratnarajah, S. Ghosh, S. Kumar, P. Chiniya, and D. Manocha, ``{AV-RIR}: audio-visual room impulse response estimation,'' *Proceedings of the IEEE/CVF Conference on Computer Vision and Pattern Recognition*, pp. 27164--27175, 2024.
\bibitem{majumder2022few} S. Majumder, C. Chen, Z. Al-Halah, and K. Grauman, ``Few-shot audio-visual learning of environment acoustics,'' *Advances in Neural Information Processing Systems*, vol. 35, pp. 2522--2536, 2022.
\bibitem{chowdhury2023adverb} S. Chowdhury, S. Ghosh, S. Dasgupta, A. Ratnarajah, U. Tyagi, and D. Manocha, ``{AdVerb}: visually guided audio dereverberation,'' *Proceedings of the IEEE/CVF International Conference on Computer Vision*, pp. 7884--7896, 2023.
\bibitem{tan2020audio} K. Tan, Y. Xu, S. Zhang, M. Yu, and D. Yu, ``Audio-visual speech separation and dereverberation with a two-stage multimodal network,'' *IEEE Journal of Selected Topics in Signal Processing*, vol. 14, no. 3, pp. 542--553, 2020.


\bibitem{he2024multi} S. He and R. Liu, ``Multi-source spatial knowledge understanding for immersive visual text-to-speech,'' *ICASSP 2025-2025 IEEE International Conference on Acoustics, Speech and Signal Processing (ICASSP)*, pp. 1--5, 2025.


\bibitem{chen2022soundspaces} C. Chen, C. Schissler, S. Garg, P. Kobernik, A. Clegg, P. Calamia, D. Batra, P. Robinson, and K. Grauman, ``{SoundSpaces} 2.0: a simulation platform for visual-acoustic learning,'' *Advances in Neural Information Processing Systems*, vol. 35, pp. 8896--8911, 2022.

\bibitem{9632680} C.J. Steinmetz, V.K. Ithapu, and P. Calamia, ``Filtered noise shaping for time domain room impulse response estimation from reverberant 
speech,'' *2021 IEEE Workshop on Applications of Signal Processing to Audio and Acoustics (WASPAA)*, pp. 221-225, 2021.

\bibitem{devlin2018bert} J. Devlin, ``Bert: pre-training of deep bidirectional transformers for language understanding,'' *Proceedings of the 2019 conference of the North American chapter of the association for computational linguistics: human language technologies, volume 1 (long and short papers)*, pp. 4171--4186, 2019.

\bibitem{valimaki2012fifty} V. V{\"a}lim{\"a}ki, J.D. Parker, L. Savioja, J.O. Smith, and J.S. Abel, ``Fifty years of artificial reverberation,'' *IEEE Transactions on Audio, Speech, and Language Processing*, vol. 20, no. 5, pp. 1421--1448, 2012.
\bibitem{valimaki2016more} V. V{\"a}lim{\"a}ki, J. Parker, L. Savioja, J.O. Smith, and J. Abel, ``More than 50 years of artificial reverberation,'' *Audio engineering society conference: 60th international conference: dreams (dereverberation and reverberation of audio, music, and speech)*, 2016.
\bibitem{binkowski2020high} M. Bi{\'n}kowski, J. Donahue, S. Dieleman, A. Clark, E. Elsen, N. Casagrande, L. C. Cobo, and K. Simonyan, ``High fidelity speech synthesis with adversarial networks,'' in *International Conference on Learning Representations*, 2020.



\bibitem{yamamoto2020parallel} R. Yamamoto, E. Song, and J. Kim, ``Parallel {WaveGAN}: a fast waveform generation model based on generative adversarial networks with multi-resolution spectrogram,'' *ICASSP 2020-2020 IEEE International Conference on Acoustics, Speech and Signal Processing (ICASSP)*, pp. 6199--6203, 2020.
\bibitem{bai2023qwen} J. Bai, S. Bai, S. Yang, S. Wang, S. Tan, P. Wang, J. Lin, C. Zhou, and J. Zhou, ``Qwen-{VL}: a versatile vision-language model for understanding, localization, text reading, and beyond,'' *arXiv preprint arXiv:2308.12966*, vol. 1, no. 2, pp. 3, 2023.
\bibitem{veaux2013voice} C. Veaux, J. Yamagishi, and S. King, ``The voice bank corpus: design, collection and data analysis of a large regional accent speech database,'' *2013 international conference oriental COCOSDA held jointly with 2013 conference on Asian spoken language research and evaluation (O-COCOSDA/CASLRE)*, pp. 1--4, 2013.
\bibitem{AMI} J. Carletta, S. Ashby, S. Bourban, M. Flynn, M. Guillemot, T. Hain, J. Kadlec, V. Karaiskos, W. Kraaij, M. Kronenthal et al., “The AMI meeting corpus: A pre-announcement,” in Proc. Int. Workshop Mach. Learn. Multimodal Interact., 2005, pp. 28–39.
\bibitem{chen2020soundspaces} C. Chen, U. Jain, C. Schissler, S.V.A. Gari, Z. Al-Halah, V.K. Ithapu, P. Robinson, and K. Grauman, ``{SoundSpaces}: audio-visual navigation in 3d environments,'' *ECCV*, 2020.

\bibitem{wenet2.0}B. Zhang, D. Wu, Z. Peng, X. Song, Z. Yao, H. Lv, L. Xie, C. Yang, F. Pan, and J. Niu, “WeNet 2.0: More productive end-to-end speech recognition toolkit,” in Proc. Interspeech, 2022, pp. 1661–1665.

\end{thebibliography}
\end{document}